\documentclass{llncs}

\usepackage{graphicx}
\usepackage{url}
\usepackage{color}
\usepackage{hyperref}
\usepackage{wrapfig}
\usepackage{lscape}
\usepackage{rotating}
\usepackage{booktabs} 
\usepackage{makecell}

\usepackage[sectionbib]{natbib}

\newcommand{\q}[1]{\lq\lq{}{}#1\rq\rq{}{}}
\newcommand{\sq}[1]{\lq{}#1\rq{}}

\begin{document}

\title{
A Workflow Model for Holistic Data Management and Semantic Interoperability in\\Quantitative Archival Research
}

\author{
    Pavlos Fafalios\inst{1} \and  
	Yannis Marketakis\inst{1} \and
	Anastasia Axaridou\inst{1} \and
	\\Yannis Tzitzikas\inst{1,2} \and 
	Martin Doerr\inst{1} 
}

\institute{
    Information Systems Laboratory, FORTH-ICS, Heraklion, Greece \and 
	Computer Science Department, University of Crete, Heraklion, Greece\\
\email{\{fafalios, marketak, axaridou, tzitzik, martin\}@ics.forth.gr}}

\maketitle            

\pagestyle{plain}

\setcounter{footnote}{0}

\vspace{3mm}

\begin{abstract}
Archival research is a complicated task that involves several diverse activities for the extraction of evidence and knowledge from a set of archival documents. The involved activities are usually unconnected, in terms of data connection and flow, making difficult their recursive revision and execution, as well as the inspection of provenance information at data element level.
This paper proposes a workflow model for holistic data management in archival research; from transcribing and documenting a set of archival documents, to curating the transcribed data, integrating it to a rich semantic network (knowledge graph), and then exploring the integrated data quantitatively.
The workflow is provenance-aware, highly-recursive and focuses on semantic interoperability, aiming at the production of sustainable data of high value and long-term validity.
We provide implementation details for each step of the workflow and present its application in maritime history research. We also discuss relevant quality aspects and lessons learned from its application in a real context.
\end{abstract}

\vspace{4mm}

\section{Introduction}

Archival research is a type of research which involves investigating and extracting evidence from archival records usually held in libraries, museums or other organisations. In its most classic sense, archival research involves the study of historical documents, thus it lies at the heart of original historical research~\citep{ventresca2017archival}. 

A large body of research in the field concerns the study of archival documents that have a \textit{repetitive} structure, such as registers, logbooks, payrolls, censuses, etc., and which provide information about one or more \textit{types of entities}, such as persons, locations, objects, organisations, etc. 
Research in this case usually starts by first collecting a set of archival documents related to a domain of interest, which are then transcribed and curated for enabling quantitative (but also qualitative) analysis of empirical facts, their description and interpretation of possible causes, influences and evolution trends~\citep{petrakis2020digitizing}.

Common data management problems in this context include: What data to transcribe and how? How to curate the transcribed data for enabling valid quantitative analysis and more effective exploration services? How to integrate the data under a common schema/model for supporting the investigation of information needs that require combining data from more than one source? How to support the long-term preservation and reuse of the data? How to maintain all data provenance information, which is important for the verification and the long-term validity of research findings that use the data? 

Consider, for instance, the real use case of the SeaLiT project\footnote{\url{https://sealitproject.eu/}} (ERC Starting Grant in the field of \textit{maritime history}), which studies the transition from sail to steam navigation and its effects on seafaring populations in the Mediterranean and the Black Sea (1850s-1920s) \citep{delis2020seafaring}. Historians in this project have collected and studied a large number of archival documents of different types and languages, such as crew lists, payrolls, and sailor registers, gathered from multiple authorities in five countries. 
Complementary information about the same entity of interest, such as a ship or a sailor, may exist in different archival documents. For example, for the same ship, one source (\textit{accounts book}) may provide information about its owners, another source (\textit{naval ship register list}) may provide construction details and characteristics of the ship (length, tonnage, horsepower, etc.), while other sources (\textit{crew lists}) may provide information about the ship's voyages and crew. 
There might also be another source (\textit{civil register}) that provides additional information about the crew members, such as their marital status and previous professions.
Data integration is very important in this context, for supporting historians in finding answers to questions that require combining information from more than one source, such as \q{finding the nationality of sailors of large ships that arrived at a specific port}. 

In addition, the name of the same entity (e.g. of a person) might be different in different sources due to typos, different language, unrecognisable characters, or use of abbreviation (e.g. \sq{G. Schiaffino}, \sq{Gaetano Schiaffino}, \sq{Gaetano Schiafino}).
Moreover, the same term, such as a profession or a ship type, may appear under different names in different sources (e.g. \sq{brigantine}, \sq{brigantino}). 
Data curation, in particular entity (instance) matching and term alignment, is crucial in this context for enabling valid quantitative analysis (like grouping a list of retrieved sailors by profession). However, at the same time, such curation must not alter the original transcribed data since this is important for verification and thus the long-term validity of the research findings.  

To cope with these problems, in this paper we describe a workflow model for holistic data management in archival research (depicted in Fig.~\ref{fig:workflow}). 
The workflow relies on the strong collaboration between researchers (domain experts) and data engineers (modeling experts), and focuses on \textit{semantic interoperability}, the ability of computer systems to exchange data with unambiguous/shared meaning \citep{ouksel1999semantic}, because such an approach supports the production of sustainable data of high value that can be extended and re-used beyond a particular research activity or project.

The workflow was designed based on real users' needs and is provenance-aware, in the sense that it retains the full provenance chain of each piece of data. It achieves this by decoupling data entry from data curation and integration.  The researcher can go back to the transcript or the original source and inspect the initial form of a piece of information. 
It is also highly-recursive, supporting the revision of the transcription, curation and integration steps, e.g. due to new knowledge acquired in the course of research. 
In comparison to related work, we treat the relevant activities in an holistic manner, paying particular attention on maintaining the provenance information at micro (data element) level, which is important for reproducible research in the age of Open Science \citep{vicente2018open}.

We showcase an implementation of the workflow model in a real use case in the field of maritime history and report empirical results from its application for satisfying real information needs of a large group of historians.
We also discuss relevant data quality aspects and lessons learned.

The rest of this paper is organised as follows:
Section~\ref{sec:rw} provides the required background and describes related work. 
Section~\ref{sec:workflow} provides an overview and the main characteristics of the proposed workflow model. 
Section~\ref{sec:workflowImpl} details how each step of the workflow model can be realised. 
Section~\ref{sec:automation} provides information about the automation of the workflow.
Section~\ref{sec:usecase} describes a real use case. 
Section~\ref{sec:quality} discusses quality aspects and relevant lessons learned.
Finally, Section~\ref{sec:conclusion} concludes the paper and outlines future work.

\section{Background and Related Work}
\label{sec:rw}

We first explain the basic notions about semantic technologies (Section~\ref{sec:notions}) and review how such technologies are used in humanities research, a large part of which concerns archival research (Section~\ref{subsec:RW_SemWeb}).
We then focus on the different data management activities towards semantic interoperability in archival research and present relevant works (Section~\ref{subsec:RW_Steps}). 
Finally, we position our work (Section~\ref{subsec:RW_Positioning}).  

\subsection{Basic Notions}
\label{sec:notions}

Semantic technologies aim at helping machines understanding data. RDF (Resource Description Framework)\footnote{\url{https://www.w3.org/TR/rdf11-concepts/}} and OWL (Web Ontology Language)\footnote{\url{https://www.w3.org/TR/owl2-overview/}} are key semantic technologies that enable encoding the semantics of data, thus allowing to formally represent the meaning involved in information \citep{antoniou2004semantic}. 
This representation has the form of a \textit{semantic network} (or \textit{knowledge graph}) which stores interlinked descriptions of \q{entities} (objects, persons, events, concepts, etc.) in a graph structure in which vertices represent entities and edges represent semantic relations between the entities. Typical standardized semantic networks are expressed as RDF triples (statements of the form \textit{subject-predicate-object}) stored in a semantic repository (RDF triplestore) \citep{ali2021survey}. 
Semantic technologies help achieving semantic interoperability, the ability of computer systems to exchange data with unambiguous/shared meaning, which is a requirement to enable machine computable logic, inferencing, knowledge discovery, and data federation between information systems \citep{ouksel1999semantic}.

\subsection{Semantic Technologies for Humanities Research}
\label{subsec:RW_SemWeb}

There is an increasing adoption of semantic technologies in the humanities field, with a main focus on how to produce and make publicly available interoperable \textit{Linked Data} \citep{heath2011linked} that can be easily queried and integrated with other datasets \citep{hyvonen2020using,hyvonen2014linked,hawkins2021archives,beretta2021challenge,fafalios2021towards}.  

\citet{oldman2016zen} provide a critical discussion on how semantic technologies and the idea of Linked Data are used in humanities research, and describe strategies for the wider adoption of these technologies for supporting high-quality digital humanities projects and the production of data that better represents human knowledge and better reflects the needs of humanities researchers.
\citet{hawkins2021archives} examines how Linked Data about archives 
is beneficial for those engaged in digital humanities research and scholarship, considering  some of the barriers that currently prevent digital humanists from being able to utilise digitised and born-digital archives. 

We believe that the workflow model that we propose, in particular its provenance-awareness at data element level, is a first step towards tackling some of the major issues described in the aforementioned works, such as the ability \q{to trace the provenance of knowledge back to the source micro‐level (with its original context and perspective intact)} \citep[p.10]{oldman2016zen}, or \q{preventing the decontextualisation and loss of nuance of archives} \citep[p. 11]{hawkins2021archives}.

With respect to historical research, for which archival research is a core part, \citet{merono2015semantic} survey the joint work of historians and computer scientists in the use of semantic technologies. 
The article provides an extensive analysis on works and systems for knowledge modelling, text processing and mining, search and retrieval, and data integration. It also discusses aspects of semantic technologies that could be furtherly exploited in historical research. Such an aspect is the \q{non-destructive data transformations} \citep[p. 22]{merono2015semantic}. Decoupling data entry from data curation and transformation, and maintaining a recursive workflow between these processes, are core characteristics of the proposed workflow model that help towards this direction.

\subsection{Data Management for Semantic Interoperability in Archival Research}
\label{subsec:RW_Steps}

Common data management activities for enabling semantic interoperability in archival research include: 
\begin{itemize}
    \item digitization / transcription of archival documents (scanning of documents, text recognition, manual transcription)
    \item documentation / metadata recording (what is the origin of a document, what is the document about, who makes the transcription, etc.)
    \item data curation / preparing the data for statistical analysis (correction or normalisation of data values, instance matching, term alignment, etc.)
    \item data integration under a common representation language (ontology-based modeling, creation of mappings, data transformation)
    \item data publication (e.g. as Linked Data)
    \item data analysis and exploration (qualitative and/or quantitative analysis, query building, data visualisation, etc.)
\end{itemize}
There is a plethora of software tools and systems for each of these activities. Below we present relevant works that have a focus on humanities research. 

\vspace{1mm} \noindent
\textbf{Digitization/Transcription.} 
One can either use text recognition software for automatically extracting text from historical documents, or manually perform the transcription process, each approach having its pros and cons. For example, the automated approach usually needs large amounts of training data and its effectiveness (quality of results) highly depends on the kind/quality of text to be extracted and the amount of training data. On the other hand, manual transcription provides high quality results but it requires a lot of effort. A mixed method is to combine automated extraction with manual correction and data entry. 
Regarding software tools, Transkribus \citep{kahle2017transkribus} is a popular platform for the digitisation of historical documents, offering AI-powered text recognition. 
FastCat \citep{fafalios2021fast} is a web application for manual and collaborative transcription based on templates. It organises the data (and metadata) in tabular forms (tables), similar to spreadsheets, offering a fast and user-friendly way to data entry.

\vspace{1mm} \noindent
\textbf{Documentation / metadata recording.} 
There are two main approaches for documentation towards semantic interoperability: 
a)~decoupling the documentation process from the ontology-based integration and the production of the semantic network, 
b)~creating the semantic network from the very beginning, i.e. during the documentation process. 
Synthesis~\citep{fafalios2021towards} is a web-based system that applies the first approach for the collaborative and scientific documentation of cultural entities (objects, events, persons, organisations, etc.), offering embedded processes for transforming the data to an ontology-based RDF dataset. ResearchSpace \citep{oldman2018reshaping} and WissKi \citep{scholz2012wisski} are platforms that apply the second approach, supporting the direct ontological representation of (meta)data. 

Spreadsheet software, such as  Microsoft Excel, and relational database management systems (RDBMS), like Microsoft Access, are still popular (and probably the dominant) tools for (meta)data entry and analysis, and are extensively used for manual documentation and metadata recording.  
There are also RDBMS-based systems, such as HEURIST\footnote{\url{http://heuristnetwork.org/}}
and nodegoat\footnote{\url{https://nodegoat.net/}}, 
that are tailored to humanities researchers and which combine a set of functionalities for building and managing research datasets, without however focusing on semantic interoperability.

\vspace{1mm} \noindent
\textbf{Data curation.} This is an optional step which is usually undertaken when a quantitative (statistical) analysis of the transcribed data is needed. In such a case, curation is very important because data quality can affect the reliability of the analysis results.  OpenRefine\footnote{\url{https://openrefine.org/}} is a popular desktop application for data cleaning. It operates on rows of data which have cells under columns (similar to relational tables). 
Silk\footnote{\url{http://silkframework.org/}} \citep{volz2009silk} is an open source framework for finding links between related data items, e.g. for instance matching. It provides a declarative language for specifying linkage rules and support of RDF link generation, through \textit{owl:sameAs} or other types of links.
For fully-automated instance matching (entity resolution), there is a plethora of learning-based methods that require manually or automatically generated training data \citep{christophides2020overview}.
Finally, the FastCat system \citep{fafalios2021fast} offers a web-based environment, called FastCat Team, which supports both automated (rule-based) and manual instance matching and vocabulary curation processes. The applied curation does not alter the original (transcribed) data and maintains links from the curated to the original data. 

\vspace{1mm} \noindent
\textbf{Data integration.} 
The objective here is to semantically represent all data and metadata using a domain (formal) ontology (as the common representation language), in order to enable semantic interoperability and make the data exploitable beyond a particular research problem or project. This activity includes the \textit{data modeling} and \textit{data transformation} processes. Data modeling consists of defining or selecting the domain ontology and creating the schema mappings, while data transformation transforms the data based on the schema mappings and creates the semantic network of integrated data.

Regarding software systems, Protégé is a popular ontology editor which provides a graphic user interface to define ontologies. It can be used for creating a new ontology for a given domain in OWL, 
or for building an extension of an existing ontology. 
For the creation and execution of schema mappings, R2RML\footnote{\url{https://www.w3.org/TR/r2rml/}} is a W3C standard for mapping relational databases into RDF, while \citet{dimou2014rml} describe an extension called RML for mapping heterogeneous sources into RDF. 
Finally, the X3ML toolkit \citep{marketakis2017x3ml} provides a declarative (XML-based) mapping definition language as well as a set of tools for the creation and maintenance of the schema mappings, and the actual transformation of the data to RDF.

\vspace{1mm} \noindent
\textbf{Data publication} 
The integrated data can be now imported in a semantic repository (RDF triplestore), either publicly available or private, which offers an Application Programming Interface (API) for accessing the data and running structured queries using the SPARQL\footnote{\url{https://www.w3.org/TR/sparql11-overview/}} protocol and language. Then, user-friendly applications can be built on top of this API for supporting end users in exploring and analysing the integrated data.  
The data can be also published as Linked Data, following the Linked Open Data (LOD) principles \citep{heath2011linked}. 
The Sampo model\footnote{\url{https://seco.cs.aalto.fi/applications/sampo/}} \citep{hyvonen2014linked} provide a framework for collaborative publishing and using of LOD, which has been tested in several domains by building the so-called \lq{}Sampo portals\rq{}~\citep{hyvonen2020sampo}.

\vspace{1mm} \noindent
\textbf{Data exploration and analysis.} 
There are two main general methods that can be used for exploring the integrated data:  
(a) \textit{free text search}: the user provides a set of keywords or a natural language question, as in ad-hoc information retrieval,
(b) \textit{interactive interface}: the user is supported by the system to express an information need, through a user-friendly interactive interface. 
In both cases the result is (usually) a ranked list of entities from which the user can start exploring relevant information, e.g. through browsing, faceted search, or different visualisations such as charts, maps, timelines, etc. 

There is a plethora of different methods for implementing keyword search over RDF data, e.g. using a document-centric information retrieval system~\citep{kadilierakis2020keyword}, or by translating a keyword query to a structured (SPARQL) query~\citep{izquierdo2021keyword}. 
For the presentation of the keyword search results, \citet{nikas2020keyword} suggest a multi-perspective approach that offers multiple presentation methods (perspectives), allowing the user to easily switch between these perspectives and thus exploit the added value of each one. 
Regarding interactive interfaces, A-Qub~\citep{kritsotakis2018assistive} and ResearchSpace~\citep{oldman2018reshaping} offer user-friendly environments which support end users in gradually building complex questions (corresponding to SPARQL queries) that associate different types of entities and information.

\subsection{Positioning}
\label{subsec:RW_Positioning}

To the best of our knowledge, there is no related work that approaches the data management part of archival research in an holistic manner, in the sense that the proposed workflow model enables the representation and efficient management 
of information, applies semantic data integration facilities in order to provide a rich knowledge graph of archival data, and at the same time it preserves the full provenance chain allowing researchers traverse from the final semantically integrated collection back to the original and transcribed manuscripts and vice versa.

\section{Workflow Model: Overview and Main Characteristics}
\label{sec:workflow}

We first provide an overview of the workflow model (Section~\ref{subsec:overview})  and then highlight its distinctive characteristics (Section~\ref{subsec:worfklowChars}).

\subsection{Roles, Input/Output and Processes}
\label{subsec:overview}
\begin{figure}[h]
    \centering
    \includegraphics[width=\textwidth]{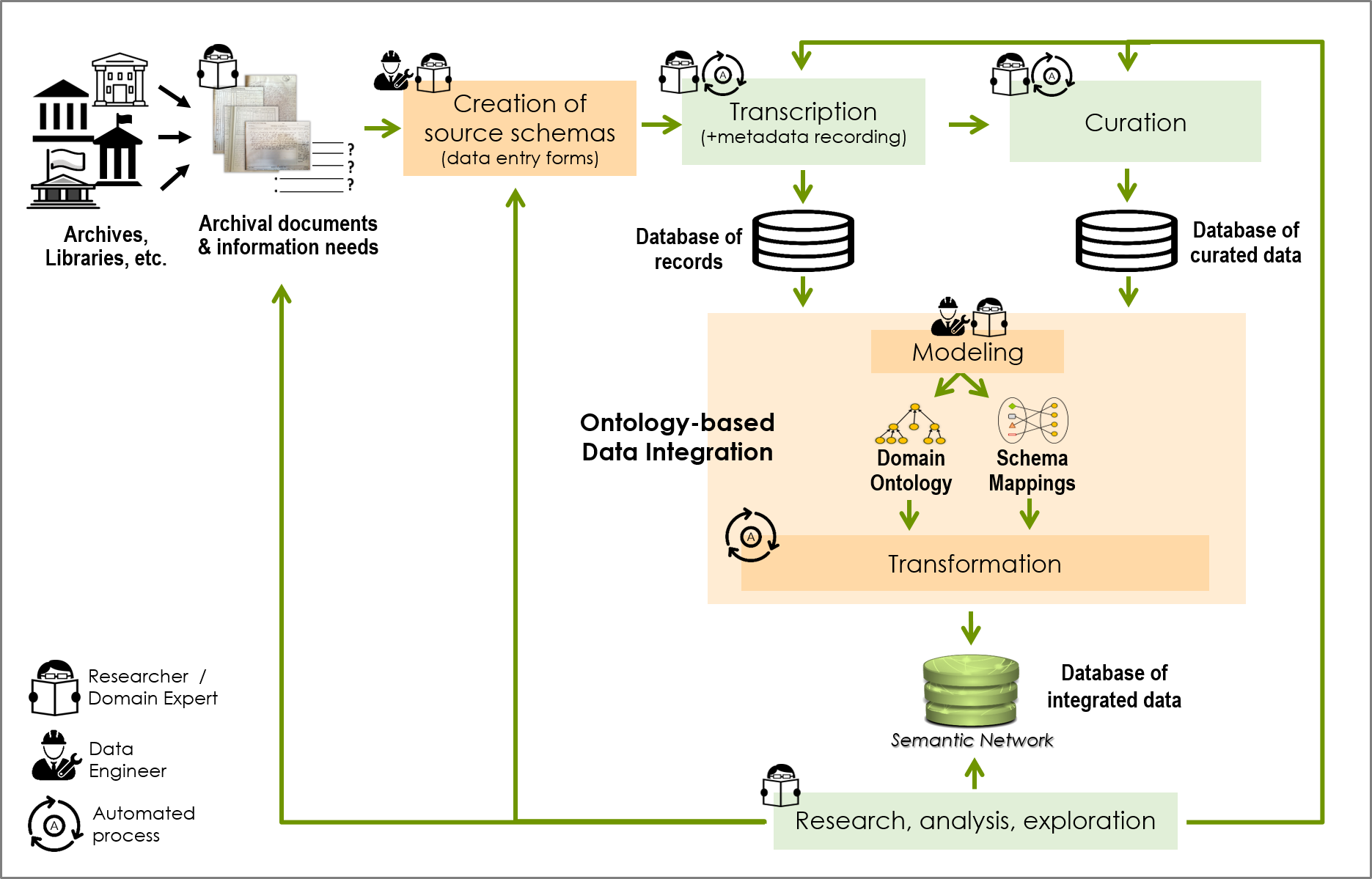}
    \vspace{-5mm}
    \caption{Workflow model for holistic data management and semantic interoperability in archival research.}
    \label{fig:workflow}
\end{figure}

Fig.~\ref{fig:workflow} depicts the proposed workflow model for supporting holistic data management in archival research.

{\bf Roles.} There are two main roles engaged in the workflow: 
a)~the \textit{researcher (domain expert / end user)}, who collects and studies the archival material, provides domain knowledge, and defines requirements, and
b)~the \textit{data engineer (modeling expert)}, who designs and implements the different workflow processes.

\vspace{2mm}
{\bf Input/Output}. The input of the workflow is a set of \textit{archival documents} gathered from different authorities by one or more researchers, together with \textit{information needs} provided by the researchers that are related to their research aims and for which the gathered archival material can provide important information (evidence). The gathering of information needs is very useful in this stage because it allows data engineers to better design and implement the next workflow processes. 
The output of the workflow is a rich semantic network (a \textit{knowledge graph}) of integrated information, which is used by the researchers for data analysis and exploration, as well as two distinct, intermediate databases: a database of records (original transcripts), and a database of curated data (curated entity instances and vocabulary terms).

\vspace{2mm}
{\bf Process 1: Creation of Source Schemas.}
Following the description logic based framework for information integration as introduced by \citet{calvanese1998description}, we first need to create the source schemas, one for each different type of source, which provide the required data entry forms in a software system for the transcription and documentation of the original archival documents.
This first step enables data curation and consolidation relative to source model semantics, as well as modeling and integration under a common ontology which can be modified in the course of research, without this affecting/delaying the transcription process. 
The close collaboration between the researchers and the data engineers is very important in this process for properly designing the schemas and avoiding mistakes during data entry that can cause difficulties/limitations in the next steps. 
An example of such a mistake is the use of a single data entry field for the recording of a measurement unit and value. This is very likely to cause issues to the end user when wanting to perform comparisons during data exploration.

The creation of a new source schema, or a modification/extension of an existing one, will be required if new archival documents of a different type of source are gathered by the researchers and need to be transcribed. This can happen at any stage of the overall pipeline and does not affect the other processes that can run in parallel for the existing gathered material.  

\vspace{2mm}
{\bf Process 2: Transcription.}
After having created one or more source schemas for the gathered archival documents, the \textit{transcription} of the documents can begin by the researchers using a software system that offers the required data entry forms. 
Apart from the transcription of the important document contents, this step includes the recording of metadata information for both the documents (archive/library, dating, etc.) and the transcription process (who makes the transcription, etc.).
The result of the transcription process is a database of transcripts. 
This is a task solely performed by the group of researchers, but which can make use of software tools for facilitating/automating transcription, such as text recognition software. 

\vspace{2mm}
{\bf Process 3: Curation.}
The next step of the workflow is the \textit{curation} of the transcribed data. At this stage researchers need to harmonise the different data elements that appear in the transcripts and resolve identity ambiguities, so that different elements that co-refer to the same real-world entity/concept receive the same identifier, and false co-references are disassociated. 

The data elements can be divided into two main categories: 
(a)~\textit{universals}; concept instances that belong to a specific vocabulary or thesaurus of terms, such as professions, object types, etc., and
(b)~\textit{particulars}; entity instances that belong to specific categories and are accompanied by characteristics/properties, such as \textit{persons} (first name, last name, birth date, etc.), \textit{locations} (name, type, etc.), \textit{organisations} (name, location, etc.).
Curation can also include the provision of corrected/preferred values (e.g. correcting the first name of a person instance) or the entity enrichment  (e.g. adding coordinates to a location instance), tasks which are usually important for better data exploitation and visualisation by the external services that operate over the curated and integrated data. 

The curation process is a task performed by the group of researchers and may include both manual and automated steps. For example, instance matching of entities, or alignment of vocabulary terms, can comprise both an automated step (based on rules) and a manual step (for validation of ambiguous cases). 
The result is a distinct database of curated data, with links  to the original data elements, which means that the curation step does not alter the data as transcribed from the original sources. 

\vspace{2mm}
{\bf Process 4: Ontology-based Data Integration.}
The next step is the ontology-based integration of the transcribed and curated data, which includes the {\em modeling} and {\em transformation} sub-processes.

For modeling, the good practice suggests to either use an established domain model (if such a model is available for the application domain), or create a new model (a specialised extension) that is compatible to an established upper ontology.  This process usually requires extensive discussions between the domain experts, who know the data, and the data engineers, who build the domain ontology and create the mappings. 

An important part of the modeling process is the creation of the \textit{schema mappings} that describe how the input data (transcripts and curated data) are mapped to classes and properties of the domain ontology. 
In general, the creation of the schema mappings can be a time-consuming process when the source schemas are many and large/complex. Nevertheless, it needs to be done only once for each different type of source, while revisions may be required if there are changes in the schemas or the target ontology. 
The use of a declarative language for defining the mappings, such as X3ML \citep{marketakis2017x3ml},  is recommended because local changes in the sources require local changes in the mapping specifications that are easy to locate and perform.

The \textit{transformation} process takes as input i) the databases (outputs of transcription and curation processes), ii) the domain ontology, and iii) the schema mappings, and produces a rich semantic network of integrated data. 
This step can be fully automated and can repeated for any new data sources that are transcribed and curated, as long as there is no change in the transcription schemas.  

\vspace{2mm}
{\bf Process 5: Research, analysis, exploration.} 
The resulting semantic network of integrated data is exploited by the researchers through one or more services that operate over the semantic network and which offer user-friendly interfaces for data browsing, analysis, and exploration. 
Here it is important for the end users to be able to go back to the transcripts, or even the scans of the original sources, for inspecting the initial form of a piece of information (before its curation and transformation), or for gathering further contextual information. 
In addition, in the course of research, a user may identify that corrections are needed in the transcribed or curated data, thus researchers need to be able to revisit the transcription and curation steps, make corrections, and then re-transform (automatically) the data for updating the semantic network. Likewise, new archival documents might be collected at any time, which means that one or more new source schemas and corresponding mappings might need to be created for enabling their transcription, curation and transformation.

\subsection{Workflow Distinctive Characteristics}
\label{subsec:worfklowChars}

Below we highlight and motivate the distinctive design and methodological characteristics of the proposed workflow model: 

\begin{itemize}
    \item \textbf{Strong collaboration between researchers (domain experts) and data engineers (modeling experts).} Such a collaboration is required for  
    a)~better designing the source schemas (and the corresponding data entry forms), 
    b)~better defining/designing the target (domain) ontology and creating the schema mappings, and
    c)~better creating/configuring the user interfaces of the data exploration service(s). 
    
    \item \textbf{Decoupling data entry from data curation and maintaining links from the curated to the original data.} This is very important not only for maintaining the data provenance, verifying information, and thus validating the research findings that make use of the data, but also because data curation and consolidation may be ambiguous and require further research and repeated revision at any time in the future (by the same or other researchers).
    
    \item \textbf{Separating source schema creation from ontology modeling.} We aim at removing the bias of the initial research hypothesis from the target (integration) model, one of the most severe philosophical problems of unbiased research and at the core of the discussion about scientific realism \citep{turner2007making,chapman2018evidential}. 
    The target model (ontology) can be developed in parallel with the data entry process and can be re-adapted at any time to new insight from the sources, without invalidating the entered data and without this affecting (or delaying) the transcription and curation processes. 
    
    \item \textbf{Separating the databases (of transcripts and curated data) from the semantic network.} Decoupling data entry and curation from the creation of the semantic network enables maintaining the semantics of the source model by keeping the transcripts as close to the original (archival) document as possible (trying to maintain their original structure), offering at the same time a familiar way to data entry that can highly speed up this time consuming process.
    In addition, this allows the straightforward production of different versions of the semantic network, considering different ontologies, or different versions of the same ontology (this only requires creating the schema mappings based on the desired target model).
\end{itemize}

\pagebreak

\section{How to Realise the Workflow}
\label{sec:workflowImpl}

We now provide implementation details for realising the workflow. 

\subsection{Faithful, Fast and Collaborative Data Transcription}

Common requirements that a data transcription system should satisfy, include: 

\begin{itemize}
    \item Supporting the \textit{faithful} and \textit{structured} transcription of information from the archival documents (as exact to the original information as possible), as well as the recording of \textit{metadata} information. 
    \item Supporting \textit{fast} data entry through an intuitive user interface that researchers are familiar with or can quickly get familiar with.
    \item Supporting the \textit{collaborative} transcription by more than one researcher, making use of the same structures (source schemas) for data entry.
\end{itemize}
These characteristics can highly affect the usability of the data entry system and thus its acceptance by the end users (researchers). 

For enabling the next \textit{data curation} process, we first need to identify what are the main entity categories (like persons, locations, objects, etc.) and the main vocabularies or hierarchies of terms that appear in the transcribed data and need curation. 
To this end, we need to define the fields in the data entry forms that provide entity or term related information. For example, the data entry fields  \textit{first name} and \textit{last name} provide information for a person instance, while the field \textit{profession} provides a vocabulary term. The values of these fields must be copied (ideally, automatically) to a new environment that allows for their curation without altering the original data as it appears in the transcripts. We then only need to provide a link from the curated to the original data and/or position information (e.g. record name, table name, row number), in order to retain the provenance information.

\subsection{Provenance-aware Data Curation}

Data curation activities that need to be supported by a dedicated software system include: 
\begin{itemize}
    \item \textit{Correcting} the name of an entity or the value of one of its properties (by setting a preferred label).
    \item \textit{Instance matching:} matching two or more entity instances that refer to the same real-world entity, which means that they must receive the same identity.
    \item \textit{Instance unmatching}: unmatching a specific entity instance from a set of automatically matched instances, which means that the instance will receive a different identity.
    \item \textit{Enrichment}: complementing an entity instance with additional information, like adding coordinates to a location.
    \item Providing a \textit{preferred term} for a vocabulary term (e.g. a term from a fixed thesaurus, or a term in English for a term in another language).
    \item Providing a \textit{broader term} for a vocabulary term (thereby creating an hierarchy of terms).
\end{itemize}

Instance matching in this context can be multi-stage. A first automated step can assign the same identity to a set of entity instances having some common characteristics, e.g. common first name, last name, and  birth date, in the case of person instances (rule-based approach), or make use of machine learning techniques (supervised or semi-supervised approach) \citep{christophides2020overview}. Then, a second manual step (performed by the researchers) can match additional entity instances that the automated step did not manage to match, or unmatch an entity instance that was incorrectly matched to other instances by the automated step. 

The instance matching/unmatching activities and the provision of preferred terms for vocabulary terms are of key importance for valid quantitative (statistical) analysis over the integrated data. Consider, for example, that a researcher who studies archival documents related to maritime history (like crew lists) wants to find the birth place of sailors that arrived at a specific port, or group them by their profession. Providing the same identity to all sailor instances that represent the same real-world person, as well as providing the same \sq{preferred} term for all different professions that correspond to the same profession, ensures that the generated aggregated information is correct.

\subsection{Ontology-based Integration}

The ontology-based integration of the transcribed and curated data consists of the below tasks:
\begin{enumerate}
    \item Data modeling using a domain ontology.
    \item Creation of schema mappings and definition of how to generate the entity identifiers (URIs).  
    \item Running the transformations for producing the semantic network of integrated data.
\end{enumerate}

\vspace{2mm} \noindent
{\bf Data modeling.}
CIDOC-CRM\footnote{\url{http://www.cidoc-crm.org/}} \citep{doerr2003cidoc} is a high-level, ISO standard ontology (ISO 21127:2014)\footnote{\url{https://www.iso.org/standard/57832.html}} of human activities, things and events happening in space and time, thus it can be used for modeling the transcribed data and supporting semantic interoperability and long-term data preservation. 
Depending on the application domain, an extension of CIDOC-CRM might be required for specialising particular notions of interest. 
For instance, in our use case we created the SeaLiT Ontology, an extension of CIDOC-CRM for the modeling and integration of data related to maritime history (more in Section~\ref{sec:usecase}).
For semantic data management using CIDOC-CRM, \citet{tzitzikas2022cidoc} analyse the relevant processes and tasks, and review the literature on applying machine learning techniques for reducing the costs related to compliance and interoperability based on CIDOC-CRM.

\vspace{2mm}  \noindent
{\bf Mapping \& Generation of Identifiers.}
This step defines how the transcribed and curated datasets will be transformed 
so that they will eventually construct the semantic network. 
The challenge is to preserve the full provenance chain, from the curated data to the original data of the 
transcript of the source, so that researchers can easily validate, further improve, or seek for further information.
The first part of this step is the definition of the schema mappings, that identify which parts from the input schema (e.g. a particular table column)
will be mapped to concrete classes and properties of the domain ontology, 
ensuring that the semantics of the original data are well-defined, non-ambiguous, and no data is lost. 
The second part defines how resource URIs and labels will be generated. 
At this point URIs will be used as the \sq{glue} connecting relevant pieces of information.

\begin{figure}
    \vspace{-1mm}
    \centering
    \fbox{\includegraphics[width=\textwidth]{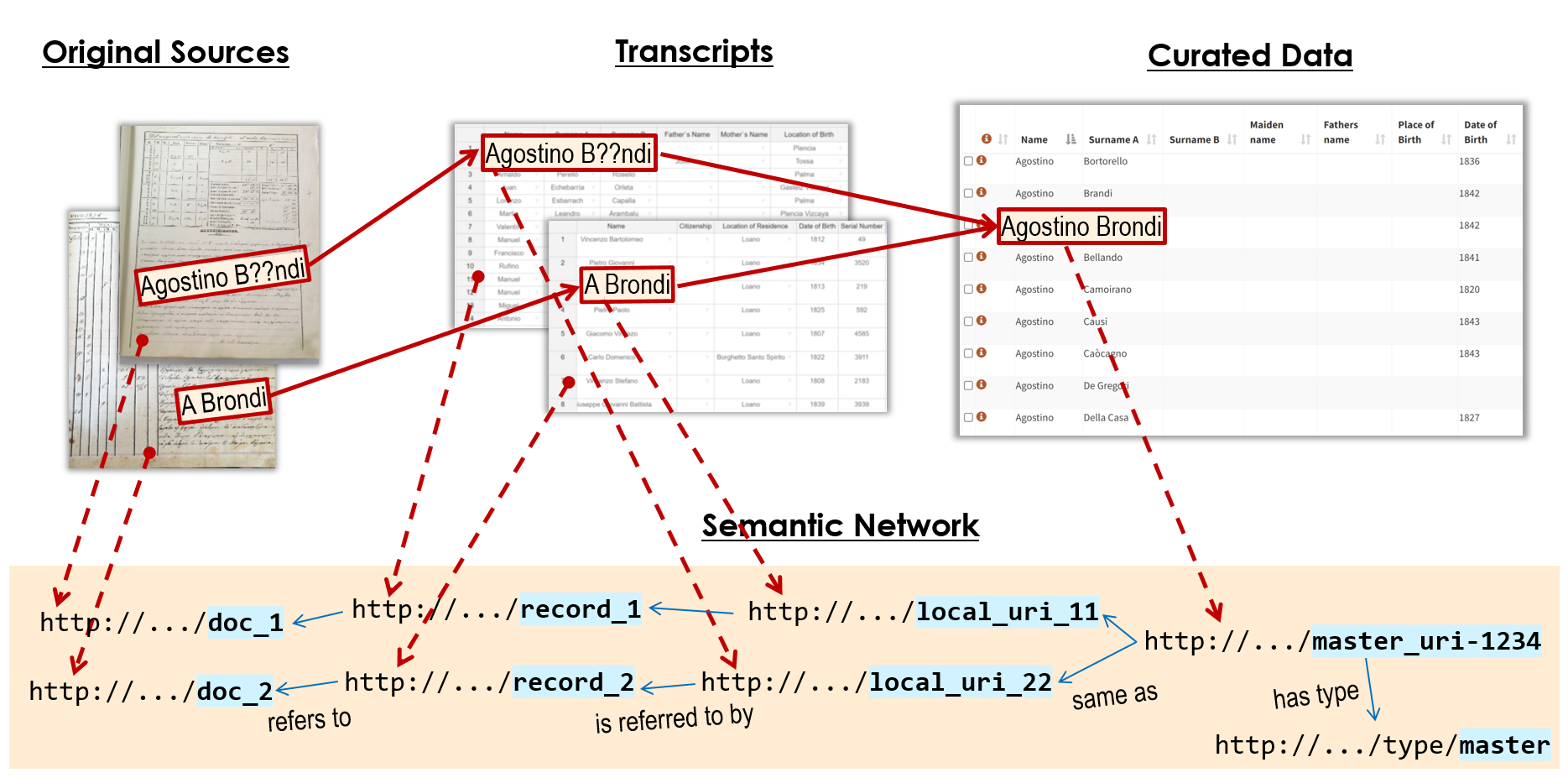}}
    \vspace{-4mm}
    \caption{Identity (URI) management and provenance chain.}
    \label{fig:provChain}
    \vspace{-1mm}
\end{figure}

Fig.~\ref{fig:provChain} shows an indicative example
on how URIs are used for establishing such connections.
In this example there are two different transcription records, each one of them describing various persons.
In one of them there is a person called `Agostino B??ndi'
(i.e. the question marks reveal that the characters could not be recognised from the original source), 
and in another one there is a person called `A Brondi'. 
For these persons two different URIs are created, since their names do not match and also they are found in different records. 
However in the curated dataset, historians agreed that these references point to the same person.
Therefore a new person instance is created, with a new URI, linked to the previous ones. 
This new instance is called \sq{master}, while the linked instances are considered \sq{local}.

Each URI consists of three parts: 
(a) the URI prefix which is common for all the resources,
(b) the type or hierarchy of the resource,
(c) the actual or hashed content of the resource.
An indicative URI is: \textit{https://rs.sealitproject.eu/kb/location/ sardinia}. 
We should also mention, that there are cases where the aforementioned strategy is not applied. An indicative case is the construction of intermediate nodes in the semantic network, for which 
a URI is not required (e.g. the \sq{E67 Birth} event). In such cases a random UUID is assigned for them.

\vspace{2mm}  \noindent
{\bf Transformation.}
This step takes as input (a) the transcribed and curated datasets and 
(b) the definitions of the schema mappings and URI generators,
and produces the ontological instances (RDF triples) with respect to the domain ontology, that are the core contents of the semantic network.
This step does not require any human intervention and can be fully automated. 
One apparent advantage of this automation is that the semantic network can be fully or partially refreshed as soon as new data have been transcribed and/or researchers have curated more data.

A good practice for managing the semantic data in terms of updating and versioning flexibility is the use of \textit{named graphs} \citep{carroll2005named}, one for each source record. 
When there is a new version of a record, or of its mapping definition file, the record output produced with a new workflow cycle can be easily integrated in the semantic repository by replacing the RDF data in the corresponding named graph. 
Also, the hierarchies of terms and locations can be effectively managed and updated in distinct named graphs, as well as the result of the \textit{materialisation} process for semantically inferred statements (the production of new RDF triples as shortcuts that represent long paths, for improving query performance).

\subsection{Semantic Network Exploitation}

The integrated data of the semantic network can be now exploited as a primary source for archival research. This includes finding answers to complex information needs and analytical queries that require combining information from different sources, as well as visualising the results in different forms, such as tables, charts, timelines, or maps, for direct use in research. 

The actual information needs depend on the application domain and the type of exploration or analysis needed by the end users.
The challenge here is to provide researchers with user-friendly and intuitive-to-use interfaces that they can trust for expressing their information needs and findings relevant information. 
Thus, the key success factors of such data exploration services are usability and trustworthiness. The latter can be achieved by enabling users to directly inspect the provenance of the displayed information, by allowing them to directly visit the transcript containing the information, or even a scan of the original archival document.
 
Some general categories of information needs include: 
(i)~finding information about a particular entity, such as the birth date and place of a person;
(ii)~retrieving a list of entities based on one or more properties of these entities (e.g. all persons having a specific residence location);
(iii)~grouping a list of retrieved entities based on some property or characteristic (e.g. grouping all retrieved persons by their profession);
(iv) finding comparative information related to some entities (e.g. number of persons employed by the organisation in different time periods).

Finally, a strategy on how to handle missing values in the data, which is very common for certain types of archival documents, is very important in order to get valid aggregated information and make safe conclusions. For example, the residence location for some persons might be empty in the original document. When grouping a set of persons by their residence location, there must be an \sq{unknown} value for this missing information.

\section{Workflow Automation}
\label{sec:automation}

The systems used in the transcription and curation processes need to intercommunicate for automating the copy of the data elements (entities, terms) that need curation from the transcription system to the curation system.
Then, an important part of the workflow can be fully automated as long as the modeling process has been completed and the mappings for all different source schemas have been created (tasks that need to be done \textit{once} for each different type of archival documents). In this case, new transcribed and curated data can be automatically transformed and imported in the semantic repository of integrated data, and thus directly be explored by the end users through the data exploration application. 

Specifically, the workflow scenario is the following: a group of researchers have collected a first set of archival documents and the data entry forms have been created in a dedicated system for each different type of source. 
The researchers start the transcription process. When transcription has been completed for the collected set of archival documents, the data elements that need curation are automatically copied to the curation environment and researchers start curating them.  
At the same time, data engineers, with the support (domain knowledge) of the researchers and by studying the available material evidence and the experts' requirements, define the target (domain) ontology and create the schema mappings for each different type of source. 
When both the transcription and curation processes have been completed for all (or a large set) of the archival documents, and the corresponding schema mappings have been created, researchers can \sq{publish} the data, which means that the transformation process is executed and the semantic network is created and ingested in a semantic repository.
Researchers can then start exploring the integrated data through the user-friendly interface of an application that operates over the semantic repository. 

At any time, researchers can transcribe and curate new archival documents, or make corrections in the existing (curated) data due to new knowledge acquired in the course of research, and then re-execute the transformation process and update the semantic repository automatically. The changes in the semantic repository are directly (and automatically) reflected in the data exploration application. 

The entire set of archival documents to be considered by the researchers does not need to be known from the beginning, meaning that new documents might be collected for transcription at any time. In this case, creation of new source schemas (data entry forms) is needed if such new documents belong to a new type of source which is different from the existing ones. 
Accordingly, changes in an existing data entry form might be needed (e.g. addition of a new column) in order to enable the transcription of a new important type of information that was not originally planned or known for an existing type of source.
In both cases, revision/extension of the domain ontology might be needed, as well as creating new schema mappings or applying changes in the existing ones.

Note here that, even if there are changes in the transcription schemas and the integration model, which actually occur during the course of a project, such changes are independent of the other transcription and curation processes performed (in parallel) by the researchers (thus, they do not affect or delay them). Moreover, the full automation of the data transformation step reduces the overhead for the researchers to the absolute minimum.

The two steps of the workflow that are the most time consuming are the \textit{transcription} and \textit{curation} processes. As already stated, several sub-tasks in these two processes can be automated or semi-automated, e.g. using state-of-the-art text recognition software \citep{kahle2017transkribus}, or applying automated instance matching / entity resolution \citep{christophides2020overview}. Here the challenge is to find the best trade-off between fully automating the tasks and having  results of high accuracy for enabling valid data analysis. 
We suggest semi-automated solutions that consider human-in-the-loop for ensuring high quality~\citep{wu2022survey,gurajada2019learning}.

\section{Use Case in Maritime History Research}
\label{sec:usecase}

The workflow has been fully implemented in a real use case for supporting a large number of historians in managing a diverse set of archival sources related to \textit{maritime history}. The context is the project SeaLiT\footnote{\url{https://sealitproject.eu/}}, in which maritime historians study the transition from sail to steam navigation and its effects on seafaring populations in the Mediterranean and the Black Sea (1850s-1920s).

Below we provide details on how each process of the workflow was implemented and illustrate an example on how a real information need provided by the historians is satisfied by exploiting the integrated data.

\vspace{2mm} \noindent
{\bf Archival material and information needs.} 
The archival material studied in SeaLiT covers a variety of sources in five languages (Spanish, Italian, French, Russian, Greek), including crew and displacement lists, registers of different types (sailors, naval ships, students, etc.), logbooks, payrolls, account books, employments records, and censuses. Details about the full archival corpus and its origin is a available in the project’s web site.\footnote{\url{https://sealitproject.eu/archival-corpus}}

Our first task was to gather a set of information needs from the historians of SeaLiT, related to their research aims and for which the studied archival material can provide important information. This is fundamental for better designing the source schemas (data entry forms), the integration model, as well as the data exploration services.
We collected around 100 information needs. Indicative examples are:\footnote{The full list of gathered information needs is available at \url{https://users.ics.forth.gr/~fafalios/SeaLiT_Competency_Questions_InfoNeeds.pdf}}  
\begin{itemize}
    \item What are the places of construction of ships during a specific period?
    \item What are the most popular European destinations (under a chronological perspective) of the ships from the Black Sea?
    \item How many people that arrived at a specific place (e.g. Barcelona) have place of birth more than X miles away? 
    \item How many ship owners per ship during a specific period?
\end{itemize}

\vspace{2mm} \noindent
{\bf Creation of source schemas and transcription.} 
The FastCat system~\citep{fafalios2021fast}, which is available as open source software\footnote{\label{foot:fastcat}\url{https://github.com/isl/FastCat}}, was used for the creation of the source schemas and the transcription of the archival documents by around 30 users in 5 countries (historians of SeaLiT).
In FastCat, users can transcribe documents and provide metadata information by creating \sq{records} belonging to specific \sq{templates}.
A ‘record’ organises the data and metadata of an archival document in a set of tables, while a ‘template’ represents the structure of a distinct data source, i.e. it defines the data entry tables, their columns as well as the type of each column (for denoting columns that provide vocabulary terms or entity-related information, whose values will be curated after transcription).
For the case of SeaLiT, twenty templates were created, one for each different type of archival source. Table \ref{tab:archSources} provides the templates as well as an overview of the information that can be recorded in each template.

\begin{table}
\vspace{-1mm}
\begin{center} 
\caption{Considered archival sources and overview of recorded information.}
\label{tab:archSources}
\vspace{-4mm}
\scriptsize
\begin{tabular}{p{4.2cm}|p{7.8cm}}
\toprule
\textbf{Archival source} & \textbf{Overview of recorded information}\\
\midrule
Crew and displacement list (Roll) 
&  Information about ships, crew members, ports. 
\\ \midrule
Crew List (Ruoli di Equipaggio) 
& Information about ships, voyages, crew members, ports. 
\\ \midrule
General Spanish Crew List & Information about ships, ship owners, crew members, voyages, ports.
\\ \midrule
Sailors Register (Libro de registro de marineros) & Information about sailors (including profession and military service organisation locations) 
\\ \midrule
Register of Maritime Personnel & Information about persons (including residence location, marital status, previous profession, military service organisation locations). 
\\ \midrule
Register of Maritime Workers & Information about maritime workers, ships, captains, ports.
\\ \midrule
Seagoing Personnel & Information about persons (including marital status, profession, end of service reasons), ships, destinations.
\\ \midrule
Naval Ship Register List & Information about ships (including tonnage, length, construction location, registration location) and ship owners.
\\ \midrule
List of Ships  & Information about ships (including previous names, registry port and year, construction place and year, tonnage, engine characteristics, owners).
\\ \midrule
Civil Register & Information about persons (including profession, origin location, marital status, death location and reason).
\\ \midrule
Maritime Register, La Ciotat & Information about persons, embarkation and disembarkation locations, ships, captains, patrons.
\\ \midrule
Students Register & Information about students and courses.
\\ \midrule
Census La Ciotat & Information about occupants (including nationality, marital status, religion, profession, working organisation, household role).
\\ \midrule
Census of the Russian Empire & Information about occupants (including marital status, estate, religion, native language, household role, occupation).
\\ \midrule
Payroll (of Greek Ships) & Information about ships, captains, voyages, persons, employments (including wages).
\\ \midrule
Payroll (of Russian Steam Navigation and Trading Company) & Information about ships, persons, recruitments (including salary per month).
\\ \midrule
Employment records (Shipyards of Messageries Maritimes, La Ciotat) & Information about workers (including marital status, profession, status of service in company).
\\ \midrule
Logbook  & Information about ships, captains, ports, route movements, voyage events.
\\ \midrule
Accounts Book  & Information about ships, voyages, captains, ports, transactions. 
\\
\\ \midrule
Notarial deeds & Information about deeds, notaries, witnesses, contracting parties, ships. 
\\
\bottomrule
\end{tabular}
\end{center}
\end{table}

\begin{figure}[h]
    \centering
    \fbox{\includegraphics[width=\textwidth]{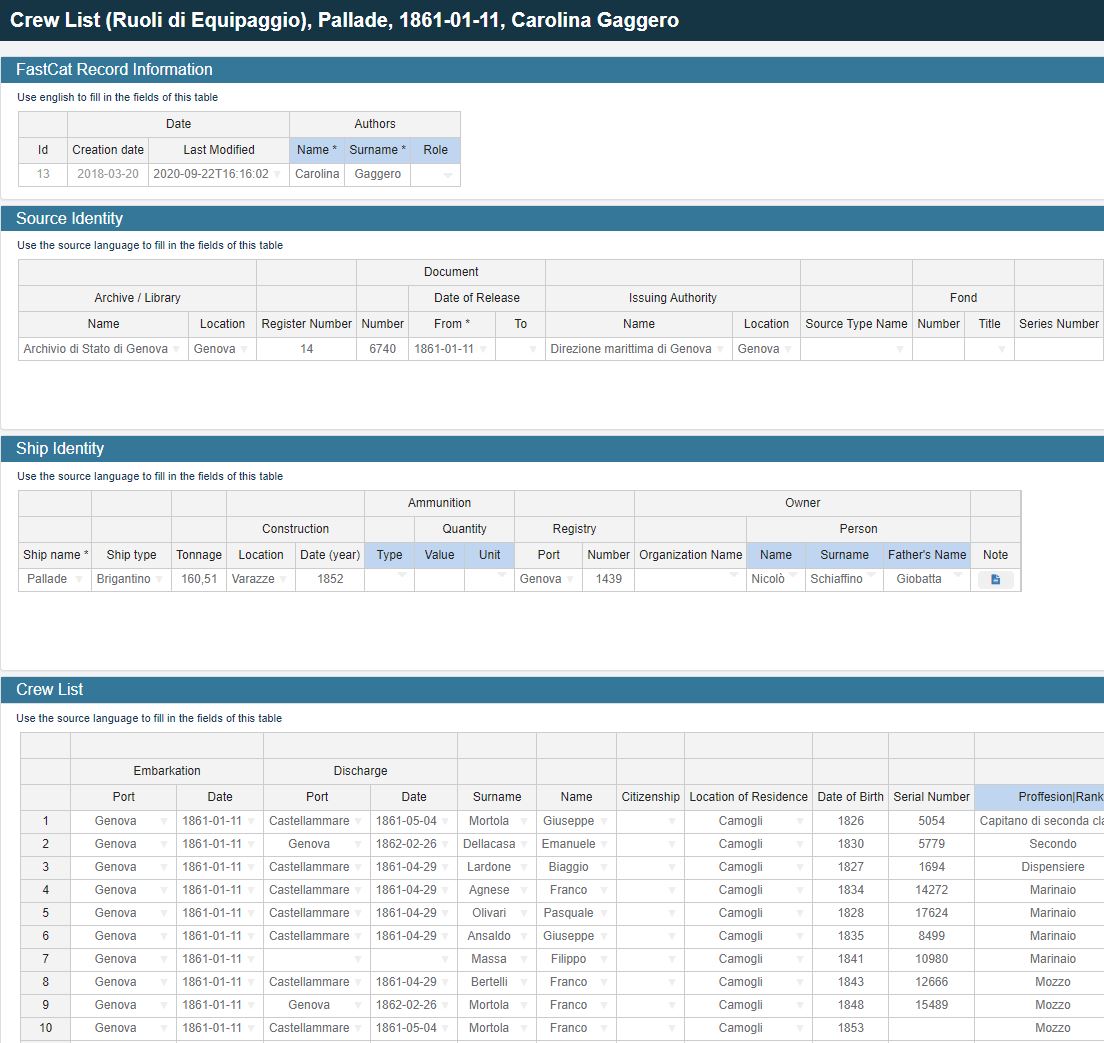}}
    \caption{An example of a real FastCat record belonging to the template \sq{Crew List (Ruoli di Equipaggio)}.}
    \label{fig:recordExample}
\end{figure}

The total number of records transcribed by the historians of SeaLiT is currently more than 620. 
Fig.~\ref{fig:recordExample} shows a part of a real record belonging to the template \textit{Crew List (Ruoli di Equipaggio)}\footnote{The full record is accessible at:  \url{https://tinyurl.com/2u35frya}} (there are totally 98 records belonging to this template). 
This template consists of six tables, enabling historians to provide/transcribe information about:
i) the record itself (creation date, last modification date, transcriber);
ii) the source (archive/library, location, register number, issuing authority, etc.);
iii) the ship (name, type, tonnage, construction location, etc.);
iv) the crew list (embarkation port and date, discharge port and date, surname, name, residence location, profession, payment information, etc.);
v) the documented navigation (date, duration, first planned destination, total crew number);
vi) the route (departure port and date, arrival port and date).
In the record of Fig.~\ref{fig:recordExample}, for instance, the transcriber has provided data for twenty six sailors and thirteen route ports that concern the navigation of the the ship \textit{Pallade} (type \textit{Brigantino}) from 11-01-1861 to 26-02-1862.

The creation and configuration of the templates in FastCat was not an \sq{one shot} process. New templates were created periodically based on new archival material gathered from the historians, or existing templates were changed several times even after the creation of records (e.g. by including additional columns in a table), for incorporating new (and important) type of information provided by particular archival documents. 

\noindent
{\bf Curation.} The curation of the transcribed data (vocabulary terms and entity instances) is performed through a dedicated environment within FastCat, called FastCat Team. 
Specifically, when a historian has completed the transcription of one or more documents (records), the record(s) can be \lq{}published\rq{}, which means that all data concerning vocabulary terms and entity instances are copied to FastCat Team for enabling their curation.

In the case of SeaLiT, the current number of vocabularies is fifty two (examples include: \textit{ship type, engine type, profession, marital status}), while the types (and current number) of entities that can be curated are \textit{ships} (about 2,400), \textit{persons} (about 99.200), \textit{locations} (about 9,800), \textit{legal entities} (about 1,100). 
For each term in a vocabulary, the user can provide a preferred term (in English) and a broader term, or inspect the records in which the term appears. 
For the curation of the entity instances, the user can correct values, select two or more instances for matching them (indicating that they represent the same real-world entity), unmatch a particular instance from a set of automatically-matched instances, or inspect the records in which the entity instance appears.
In the case of locations, the user is able to add an identifier (TGN/Geonames ID), as well as coordinates or a secondary location name (e.g. a historical name).

\begin{figure}[h]
    \centering
    \fbox{\includegraphics[width=\textwidth]{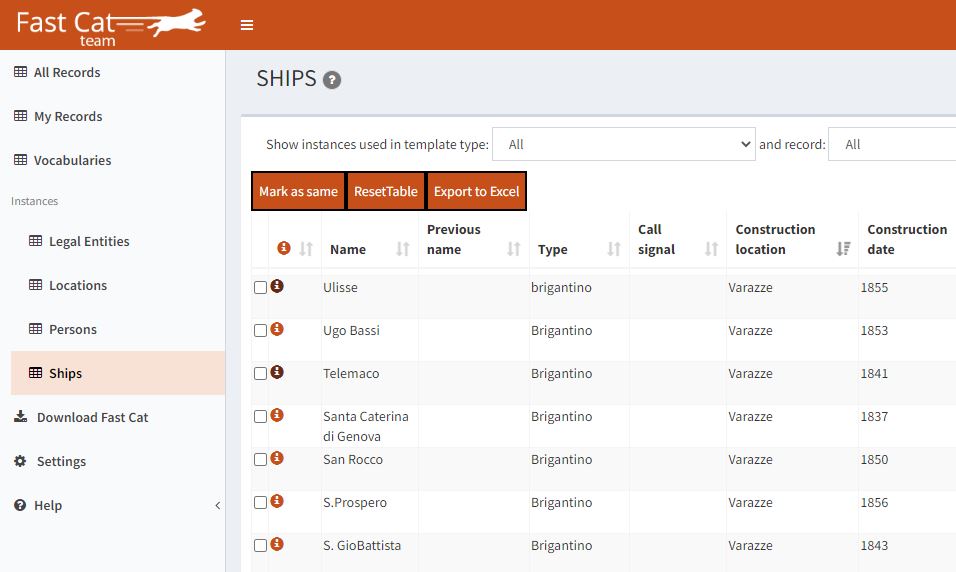}}
    \caption{Curation of ship instances in FastCat Team.}
    \label{fig:teamExample}
\end{figure}

Fig.~\ref{fig:teamExample} shows the user interface of FastCat Team, in particular the page that allows the curation of ship instances. For more information about FastCat (and FastCat Team), the reader can refer to \citet{fafalios2021fast}. 

\noindent
{\bf Ontology-based integration and transformation.} 
For data integration we created a data model compatible with CIDOC-CRM, called \sq{SeaLiT Ontology}\footnote{\url{https://zenodo.org/record/6797750}}.
The current version of the ontology (v1.1) contains forty six classes and seventy nine properties, allowing the description of information about ships, voyages, employments, payments, seafaring people, teaching courses, and other relevant activities. 
For creating the schema mappings and transforming the data to RDF we make use of the X3ML framework~\citep{marketakis2017x3ml}. In particular, one mapping definition file has been created for each template in FastCat, as well as one for each of the four categories of entities in FastCat Team and one for all the vocabularies.

The derived semantic network contains more than 18.5M RDF triples and is currently exploited by the data exploration application (ResearchSpace; more below) for supporting historians in finding answers to their information needs. The full RDF datasets are publicly available\footnote{\url{https://zenodo.org/record/6460841}}.
The network contains interconnected information for thousands of sailors, ships, locations, organisations, voyages, and many other relevant activities, as well as connections with publicly available resources (Geonames, Getty TGN).

\vspace{2mm} \noindent
{\bf Semantic network exploration.} For enabling historians of SeaLiT and other interested parties to explore the integrated data and find answers to their information needs, we make use of ResearchSpace~\citep{oldman2018reshaping}.
ResearchSpace is a configurable, open source platform which operates over a semantic network accessible through an RDF triplestore. It offers a variety of functionalities, including a \textit{query building} interface that supports users in gradually building and running complex queries through a user-friendly interface. The results can then be browsed and analysed quantitatively through different visualisations, such as bar charts.

\begin{figure}
    \centering   
    \hspace*{-1.5cm}\includegraphics[width=16cm]{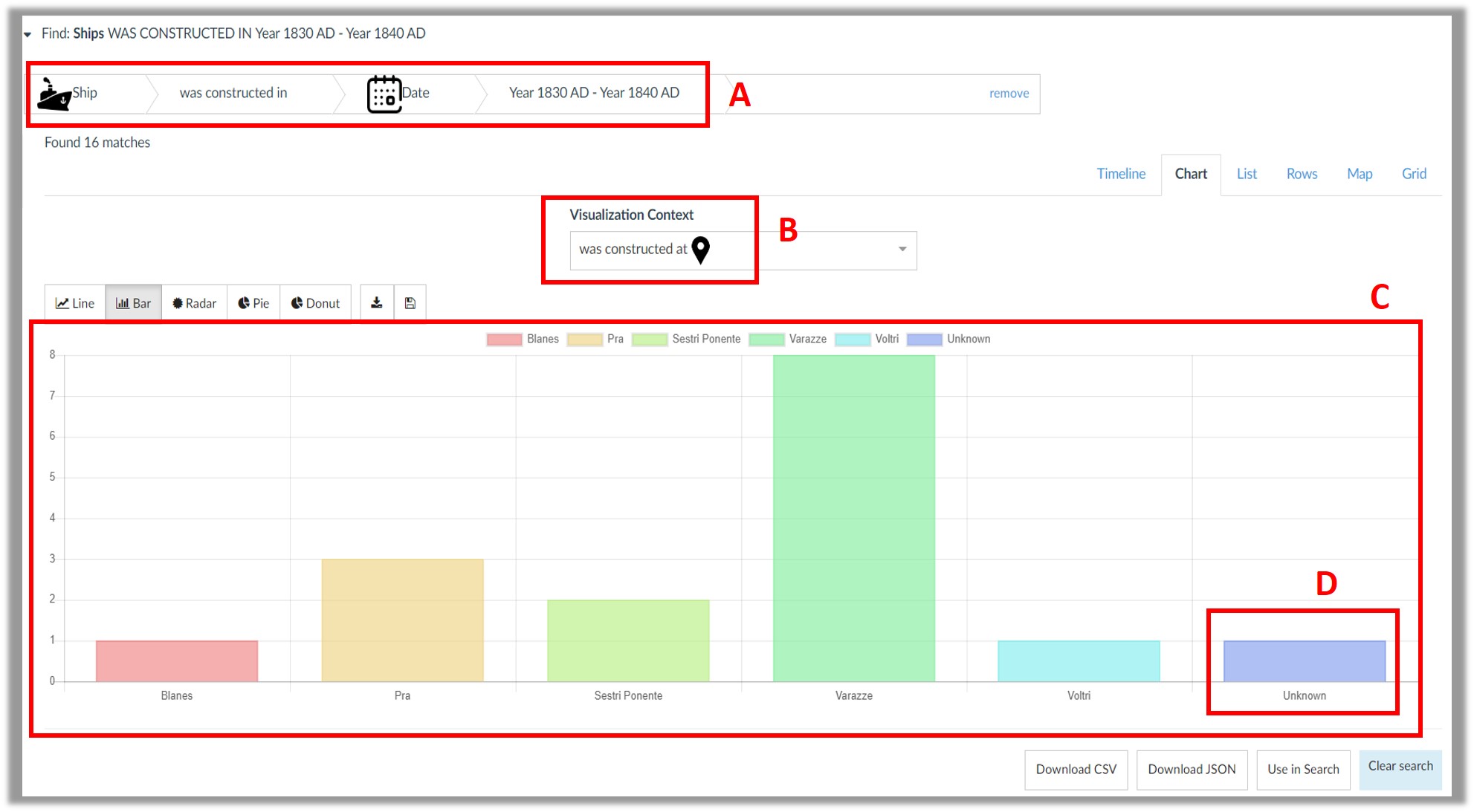}
    \caption{Semantic search and results visualisation in ResearchSpace.}
    \label{fig:RSexample}
\end{figure}

The platform was configured for the case of SeaLiT data, offering three main data exploration functionalities: 
a) keyword search, 
b) semantic search (through its assistive query building interface), and 
c) entities browsing (per type of archival source). 
Fig.~\ref{fig:RSexample} shows a screen dump of the semantic search functionality.
The user inspects the \q{construction location of ships that were constructed between 1830 and 1840}. The user first searched for ships constructed between 1830 and 1840 (Fig.\ref{fig:RSexample}-A), and then selected to group the retrieved ships by their construction location  (Fig.\ref{fig:RSexample}-B) and visualise the results in a bar chart (Fig.\ref{fig:RSexample}-C).
This query corresponds to a real information need as provided by the historians of SeaLiT, and the answer is shown to  the user instantly (in less than one second). 
If the construction location is unknown for a ship, this missing information is displayed in the chart (see \sq{Unknown} bar, Fig.\ref{fig:RSexample}-D).
The user can also start browsing information about the retrieved ships (e.g. inspecting the owners of a ship and then other ships owned by the same owner), visit the FastCat transcripts that provide the corresponding information (for validation, or inspection of additional contextual information), or download the results in CSV format for further (external) analysis. 

A deployment of the application is publicly accessible.\footnote{\url{http://rs.sealitproject.eu/}}

\section{Quality Aspects and  Lessons Learned}
\label{sec:quality}

We discuss data quality aspects as well as relevant lessons learned from the application of the proposed workflow model in maritime history research. 

\subsection{Quality Aspects}
\label{subsec:quality}
Every workflow cycle ends up with semantic data that in some cases may suffer low quality characteristics, making the data practically difficult to be exploited for the needs of research. 
In literature, data quality is commonly considered as \q{fitness for use} as well as an indicator of data usability \citep{pipino2002data,wang1996beyond}, and several dimensions and metrics for measuring data quality have been proposed \citep{pipino2002data,zaveri2016quality}. 
Although studying quality factors in detail is out of the scope of this paper, below we focus on three main quality dimensions of the semantic data that can significantly affect the quantitative analysis process: completeness, consistency, conciseness.

\textbf{Data completeness.}
A quality dimension that can be easily assessed in the context of a schema/ontology or the particular use case scenario \citep{zaveri2016quality}. The lack of essential information, like missing dates and locations of events, or names and professions of actors of a registry, may affect the research analysis and the evidence for making a decision about a historical subject.

\textbf{Data consistency.}
This dimension can be viewed from a number of perspectives \citep{zaveri2016quality,hassenstein2022data}. Our perspective comprises the \textit{schema-based} and the \textit{value-based}  (or \textit{representational}) consistency.
Schema-based consistency can be evaluated against a particular schema/ontology. It prevents modeling issues, like the incompatible attribution/interlinking of the entities, and averts potential reasoning malfunction. For example, assigning \sq{tonnage} to a person (instead of a ship) makes no sense, and under particular reasoning premises it may produce inaccurate inference that people were used for the transportation of goods. 
Value-based consistency concerns the format and the structure of comparative values (numbers, dates, measurement values) to enable comparability. Magnitudes, dimensions, quantities, time-spans, dates, places’ coordinates, etc., to be effectively compared, they have to align their reference points or units of measurement. 

\textbf{Data conciseness}. 
This quality dimension comprises two perspectives: \textit{schema-level} conciseness and \textit{instance-level} conciseness \citep{zaveri2016quality,mendes2012sieve}.
Schema-level conciseness means that the data does not 
contain equivalent attributes with different names (responsibility of the data modeling engineer), while instance-level conciseness means that the data does not contain equivalent objects with different identifiers (highly-dependant on the quality of the curation process).

\subsection{Lessons Learned}

Next we present issues related to data quality that we faced while implementing the workflow and which should be taken into account. 

\label{subsec:lessons}
{\bf Missing information. }
Missing values are very common and an important-to-know information for researchers because they can affect the accuracy of quantitative (statistical) analysis. This is related to the \textit{completeness} quality aspect described above.
When a piece of information is not provided in the original source, the corresponding cell in the data entry system is left empty. The data exploration system must consider such empty values while aggregating and showing information. 

{\bf Data entry errors.}
Errors in the transcripts during data entry are common, such as accidentally filling the wrong column in a table, or putting the information in the wrong place due to misunderstanding.  
This is related to the \textit{schema-based} \textit{consistency} quality aspect described above.
Such errors are directly reflected in the data exploration interfaces and can spoil user experience. Thus, it is important to allow researchers visit the original transcripts for validation or making corrections. Moreover, offering mechanisms in the user interface that support users to avoid such errors during data entry can limit the problem. 

{\bf Non-consistent comparative values.} It is very common that comparative values, such as dates, dimensions, quantities, location coordinates, are not consistent across archival sources of different types, because of different reference points or units of measurement, making difficult their use in comparisons, filtering, etc. 
This is related to the \textit{value-based} \textit{consistency} quality aspect described above.
An additional (automated, semi-automated or manual) step is needed for aligning such values, however without changing the values as they appear in the original source. This can happen either during data curation or during data transformation. 

{\bf Costly data curation. }
Low-quality data curation can reduce user satisfaction and produce invalid analysis results. 
This is related to the \textit{instance-level} \textit{conciseness} quality aspect described above.
The cost of manual data curation is relative to the size of the data that need curation (number of entity instances, number of vocabulary terms). The process can be very time consuming for researchers in cases such as SeaLiT where the number of entities and vocabularies is high. Thus, there is a need for tools that automate as much as possible curation without significantly affecting quality, e.g. through semi-automatic processes, supervised algorithms, or application-specific machine learning.

\section{Conclusion}
\label{sec:conclusion}

We presented a workflow model for holistic data management in archival research: from transcribing and documenting a set of archival documents, to curating the transcribed data, integrating it to a rich semantic network, and then exploring and analysing the integrated data. The merits of the approach is that it speeds up data entry, it is provenance-aware decoupling data entry from data curation and integration, it is interactive as well as appropriate for semantic interoperability, aiming at the production of sustainable data of high value and long-term validity. 

We have showcased the feasibility and effectiveness of the model in maritime history research, and we have reported empirical results from its application (about thirty users, twenty types of archival documents, more than 600 records, more than fifty vocabularies, more than 110,000 entity instances, more than 18.5 million triples of integrated information).

Issues that are worth further research include: 
(a) semi-automated methods to speedup data curation, 
(b) investigate the evolution requirements of the semantic network, as proposed by \citet{marketakis2020workflow}, 
(c) methods and interfaces to support researchers in defining and updating the source schemas by themselves.

\subsection*{Acknowledgements}
This work has received funding from the European Union's Horizon 2020 research and innovation programme under the Marie Sklodowska-Curie grant agreement No 890861 (Project ReKnow), and ii) the European Research Council (ERC) grant agreement No 714437 (Project SeaLiT).

\bibliographystyle{apalike-bold}
\bibliography{PaperX__BIB}

\end{document}